\documentclass[a4paper,11pt,twoside]{article}
\usepackage{graphicx,anst}

\begin{document}
\makeatletter
\renewcommand{\theequation}{\thesection.\arabic{equation}}
\@addtoreset{equation}{section}
\makeatother
\def\sltwo{{\sl sl}(2)}
\title{\sc\huge \sltwo\ Construction \\ of Type A ${\cal N}$-fold 
Supersymmetry}
\author{\Large Hideaki Aoyama$^{\dagger 1}$,
Noriko Nakayama$^{\star 2}$,
Masatoshi Sato$^{\ddagger 3}$\\[7pt]
\Large and Toshiaki Tanaka$^{\dagger 4}$\\[10pt]
\sc $^\dagger$Faculty of Integrated Human Studies\\[3pt]
\sc Kyoto University, Kyoto 606-8501, Japan\\[7pt]
\sc $^\star$Graduate School of Human and
Environmental Studies\\[3pt]
\sc Kyoto University, Kyoto 606-8501, Japan\\[7pt]
\sc $^\ddagger$The Institute for Solid State Physics\\[3pt]
\sc The University of Tokyo, Kashiwanoha 5-1-5, \\[3pt]
\sc Kashiwa-shi, Chiba 277-8581, Japan}
\footnotetext[1]{aoyama@phys.h.kyoto-u.ac.jp}
\footnotetext[2]{nakayama@phys.h.kyoto-u.ac.jp}
\footnotetext[3]{msato@issp.u-tokyo.ac.jp}
\footnotetext[4]{ttanaka@phys.h.kyoto-u.ac.jp}
\maketitle

\vspace{-12cm}
\rightline{KUCP-0183}
\vspace{11.5cm}
\thispagestyle{empty}
\def\nfsusy{${\cal N}$-fold supersymmetry}
\def\nfsusic{${\cal N}$-fold supersymmetric}
\def\hn{H^-_{\cal N}}
\def\thn{\widetilde H^-_{\cal N}}
\def\hnp{H^+_{\cal N}}
\def\ddq{\frac{d}{dq}}
\def\ww{\widetilde{W}(q)}
\def\wwq{\widetilde{W}}
\begin{abstract}
\nfsusy\ is an extension of the ordinary supersymmetry
in one-dimensional quantum mechanics.
One of its major property is quasi-solvability, which means
that energy eigenvalues can be obtained for a portion of the spectra.
We show that recently found Type A \nfsusy\
can be constructed by using \sltwo\ algebra, which
provides a basis for the quasi-solvability.
By this construction we find a condition
for the Type A \nfsusy\, which is less restrictive than
the condition known previously.
Several explicitly known models
are also examined in the light of this construction.
\end{abstract}
\newpage

\section{Introduction}
One novel feature of supersymmetry is its nonrenormalization theorem 
\cite{Wit,Wit2}.
The same is true \cite{AKOSW2} for higher derivative extentions of 
supersymmetry\cite{AIS}-\cite{FNN}.
In fact, this observation was crucial for 
identification of a simple form of \nfsusy,
whose supercharges of \nfsusy\ are ${\cal N}$-th order polynomials 
of the momentum.
Namely, vanishing of the leading Borel singularity
of the perturbative series in a quartic potential model
was found \cite{AKOSW} by the use of the valley method 
\cite{RR}-\cite{AKHOSW}
and this motivated a search for supersymmetry or its extension,
which lead to ``\nfsusy" in Ref.\cite{AKOSW2}. 
Later it was extended to a periodic and exponential potentials 
in Ref.\cite{ASTY} and to a sextic potential in Ref.\cite{AST}.

The (extended) nonrenormalization theorem, 
which applies to all of these models,
states that the perturbative corrections to a part of the spectrum 
are either nonexistent, or are obtained in a closed form.
When \nfsusy\ is not spontaneously broken, 
there is no nonperturbative corrections, and thus that part of
the spectra is exactly solved.  This property has also been
known in literatures, 
quite independently from the supersymmetry considerations,
as ``quasi-exact solvability" \cite{Tur}-\cite{Ush}.
When \nfsusy\ is spontaneously broken,
in general there are non-perturbative corrections to energy
eigenvalues and we only know of the perturbative part.
We dubbed this property ``quasi-perturbative solvability" \cite{AST2}.

Quasi-exact solvability is known to be connected with the
property that the Hamiltonian can be written in terms of \sltwo\ 
generators \cite{Tur}.  In Ref.\cite{AKOSW2}, we made the 
same connection for the quartic potential, which is only
quasi-perturbatively solvable.  Similar connections were made
for the sextic potential in Ref.\cite{KP2, DDT} and for exponential potential
in Ref.\cite{KP1}.
Through these works it became apparent that all of the 
models, which can be classified as {\it Type A} \cite{AST}, 
may be expressed in terms of the \sltwo\  generators.
The purpose of this letter then is to make the general connection 
between the Type A \nfsusy\ and the \sltwo\ representations.

In the following, we first review \nfsusy\ in general and
the definition of the Type A \nfsusy\ in Section 2.
The method to construct \nfsusic\ models from quasi-solvability is
briefly reviewed in Section 3.
Based on these knowledge, we 
show that quasi-solvability naturally leads to the \sltwo\ 
representations and show that this leads to Type A \nfsusy\ in Section 4.
Type A \nfsusic\ models which can be written down explicitly are examined in 
Section 5.  A summary is given in Section 6.

\section{\nfsusy\ and its Type A subclass}
We first present a concise definition of \nfsusy\ and Type A \nfsusy.
One dimensional quantum mechanical model with \nfsusy\ 
has one ordinary (bosonic) coordinate $q$ and 
fermionic coordinates $\psi$ and $\psi^\dagger$,
which satisfy the following:
\begin{eqnarray}
\{\psi,\psi\}=\{\psi^{\dagger}, \psi^{\dagger}\}=0
,\quad 
\{\psi,\psi^{\dagger}\}=1.
\label{eq:ferpro}
\end{eqnarray}
The Hamiltonian ${\bf H}_{\cal N}$ is given as follows,
\begin{eqnarray}
{\bf H}_{\cal N}=H^{-}_{{\cal N}}\psi\psi^{\dagger}
+H^{+}_{{\cal N}}\psi^{\dagger}\psi,
\end{eqnarray}
where $H^{\pm}_{\cal N}$ are ordinary Hamiltonians, 
\begin{eqnarray}
H^{\pm}_{\cal N}=\frac{1}{2}p^{2}+V^{\pm}_{\cal N}(q)
\label{eq:Hfutu}
\end{eqnarray}
with $p=-id/dq$.
The ${\cal N}$-fold supercharges are generically defined as 
\begin{eqnarray}
Q_{\cal N}=P^{\dagger}_{\cal N}\psi
,\quad
Q^{\dagger}_{\cal N}=P_{\cal N}\psi^{\dagger}, 
\end{eqnarray}
where $P_{\cal N}$ is an ${\cal N}$-th order polynomial of $p$,
\begin{eqnarray}
P_{\cal N}=
w_{\cal N}(q)\,p^{\cal N}+w_{{\cal N}-1}(q)\,p^{{\cal N}-1}+\cdots+
w_1(q)\,p+w_0(q). 
\label{sec_dons:pn}
\end{eqnarray}

The \nfsusy\ algebra is defined as follows:
\begin{eqnarray}
&&\{Q_{\cal N}, Q_{\cal N}\} =
\{Q^{\dagger}_{\cal N}, Q^{\dagger}_{\cal N}\}=0,
\label{eq:nilp}\\
&&[\,Q_{\cal N}, {\bf H}_{\cal N}\,] = 
[\,Q^{\dagger}_{\cal N}, {\bf H}_{\cal N}\,]=0.
\end{eqnarray}
Among the above, the nilpotency (\ref{eq:nilp}) is guaranteed by 
the property of the fermionic coordinates (\ref{eq:ferpro}),
while the latter leads to 
\begin{equation}
P_{\cal N}\hn-H^+_{\cal N}P_{\cal N}=0,
\label{eq:nfsusyalgebra}
\end{equation}
and its conjugate, which are essentially
${\cal N}+2$ differential equations for the
potentials $V^\pm_{\cal N}(q)$ and the 
coefficient functions $w_{{\cal N}, \cdots, 0}(q)$.
Since there are ${\cal N}+3$ functions to be determined,
one function remains arbitrary. One may, for example, choose it
one of the potentials.  Therefore it may be said that
any ordinary (purely bosonic, nonsupersymmetric) one-dimensional quantum model
may be extended so that it is a part of an \nfsusic\ system.
This does not mean that any one-dimensional system can be even partially
solved by means of nonrenormalization theorem(s) of \nfsusy:
The differential equations that are obtained from Eq.(\ref{eq:nfsusyalgebra})
are just as difficult to solve as Schr\"odinger equations.

The explicitly known models noted in the introduction
are, in contrast, solvable:
The \nfsusic\ algebra (\ref{eq:nfsusyalgebra}) can be 
solved explicitly, and the part of the spectra are solved.
All these models belong to Type A \nfsusy,
which is defined to have the following form of the supercharge \cite{AST}:
\begin{eqnarray}
P_{\cal N}=(D+i({\cal N}-1)E(q))(D+i({\cal N}-2)E(q))\cdots(D+iE(q))D,
\label{eq:typeapn}
\end{eqnarray}
where $D\equiv p-iW(q)$. 
In Refs.\cite{AST,AST2}, we showed, by induction in ${\cal N}$, 
that the algebra (\ref{eq:nfsusyalgebra}) 
is satisfied if the following conditions are met:
\begin{eqnarray}
&&V^{\pm}_{\cal N}=\frac{1}{2}\left(W^2 \pm W'\right)
\nonumber\\
&&\hspace{30pt}\mbox{}
+\frac{{\cal N}-1}{2}
\left[-EW+\frac{2{\cal N}-1}{6}E^2-\frac{{\cal N}+1}{6}E'
\pm\left(W'-\frac{\cal N}{2}E'\right)\right],
\label{eq:typeaa}\\
&&\lefteqn{(\wwq'+E\wwq)''-E(\wwq'+E\wwq)'=0}
\hspace{200pt} (\mbox{for }{\cal N}\ge 2),
\label{eq:typeab}\\
&&\lefteqn{(E'+E^2)''-E(E'+E^2)'=0}
\hspace{200pt} (\mbox{for }{\cal N}\ge 3),
\label{eq:typeac}
\end{eqnarray}
where a prime denotes a derivative with respect to $q$ and 
\begin{equation}
\widetilde{W}(q)\equiv W(q)-\frac{{\cal N}-1}{2}E(q).
\end{equation}

\section{Quasi-solvability and \nfsusy}
\nfsusy\ allows alternative construction based on 
quasi-solvability \cite{AST2}.
Since this will be a key for the \sltwo\  construction, we will
briefly review this aspect.

We first note that 
the $p^{{\cal N}+2}$-terms in the \nfsusy\ algebra (\ref{eq:nfsusyalgebra})
trivially vanish.
The $p^{{\cal N}+1}$-terms yield 
$w_{\cal N}(q)'=0$.
Therefore we choose that
\begin{equation}
w_{\cal N}(q)=1,
\label{eq:wnone}
\end{equation}
hereafter without losing generality.
Then, the $p^{{\cal N}}$-terms yield the following relation:
\begin{equation}
V^+_{\cal N}(q)=V^-_{\cal N}(q)+iw_{{\cal N}-1}(q)'.
\label{eq:vpmdet}
\end{equation}

Let us assume that an operator $P(p,q)$, whose
highest order of $p$ (when all $p$'s are moved to right of all $q$'s)
is ${\cal N}$, has a nontrivial kernel
${\cal V}\equiv\{\phi(q)\,|\,P(p,q)\phi(q)=0\}$ of dimension 
${\cal N}$.
Further let us assume that there is a Hamiltonian $H$ of the form,
\begin{equation}
H=\frac{1}{2}\,p^2+V(q)
\end{equation}
that maps the kernel ${\cal V}$ into itself:
\begin{equation}
H\phi(q)\in{\cal V} \mbox{ for any } \phi\in{\cal V}.
\label{eq:mapinto}
\end{equation}
Such a system is ``quasi-solvable" \cite{Tur}--\cite{Ush},
in the sense that the energy eigenvalues are obtained in 
a closed form for the part of the spectrum that is spanned by 
the kernel.\footnote{We do not require that the equation 
$P(p,q)\phi(q)=0$ is algebraically solvable for the system
to be "quasi-solvable".
This definition does not conflict with, e.g. the definition
of "quasi-exact solvability" in Ref.\cite{Ush}.}
This is because of the following:
Let $\phi_n(q)$ ($n=1,2,\cdots,{\cal N}$) be a basis of the kernel ${\cal V}$.
Then, the above property (\ref{eq:mapinto}) means that $H \phi_n(q)$
is given by a linear combination of $\phi_{1,..,N}(q)$:
\begin{equation}
H\,\phi_n(q) = \sum_{m=1}^{\cal N} \mathbf{S}_{n,m}\phi_m(q).
\end{equation}
Therefore, by diagonalizing the ${\cal N}\times{\cal N}$ matrix
$\mathbf{S}$, we obtain the energy eigenvalues of ${\cal N}$-states 
in a closed form.\footnote{It should be noted that we did not require
normalizability for $\phi_i(q)$ in the above.
Therefore, the normalizability of the resulting eigenfunctions should be
separately examined, especially in connection with the
perturbation theory \cite{AST2}.}

It is straightforward to show that existence of
such a Hamiltonian $H$ that satisfy 
the property (\ref{eq:mapinto}) implies that the system is \nfsusic.
This is because of the following.
Let us denote $P(p,q)$ as follows:
\begin{eqnarray}
P(p,q)=p^{\cal N}+c_{{\cal N}-1}(q)p^{{\cal N}-1}+\cdots+c_0(q),
\end{eqnarray}
where we have chosen the coefficient function of $p^{\cal N}$ 
in $P(p,q)$ equal to one, since it is irrelevant
for the definition of the kernel ${\cal V}$. (Its form
is also motivated by the allowed choice (\ref{eq:wnone}).)
We introduce another Hamiltonian $K$ as follows,
\begin{eqnarray}
K&=&\frac{1}{2}\,p^2+Y(q),\\[5pt]
Y(q)&=&V(q)+ic_{{\cal N}-1}(q)'.
\label{eq:barVdef}
\end{eqnarray}
It may be noted that the latter form is motivated by Eq.(\ref{eq:vpmdet}).
Then the operator $G(p,q)\equiv P(p,q)H-KP(p,q)$ 
contains only up to $({\cal N}-1)$-powers of $p$, by the same reason 
that lead to Eq.(\ref{eq:vpmdet}).
It also satisfies the following:
\begin{equation}
G(p,q)\phi_n=0 \quad \mbox{ for } n=1,2,\cdots,{\cal N}.
\end{equation}
Since $G(p,q)$ is an $({\cal N}-1)$-th order differential operator, it can not
non-trivially annihilate ${\cal N}$ independent functions 
$\phi_{1,\cdots,{\cal N}}(q)$. 
Therefore, the operator $G(p,q)$ is identically zero:
\begin{equation}
P(p,q)H-KP(p,q)=0.
\end{equation}
If we identify $P_{\cal N}=P$, $\hn=H$, and $\hnp=K$,
the above relation is equivalent to the \nfsusy\ algebra
(\ref{eq:nfsusyalgebra}).

The above argument shows that if we can construct 
a Hamiltonian $H$ and an operator $P(p,q)$ that satisfy
the above quasi-solvability condition 
we have a \nfsusic\ system.
In the following, we will carry out
this construction for Type A \nfsusy.

\section{\sltwo\  Construction}
We first derive a simpler expression for $P_{\cal N}$
of Type A \nfsusy\ (\ref{eq:typeapn}).
Let us first introduce
\begin{equation}
U(q)\equiv e^{\int^q W(q')dq'},
\end{equation}
and transform $P_{\cal N}$ as follows:
\begin{equation}
U P_{\cal N} U^{-1} \equiv (-i)^{\cal N}\widetilde P_{\cal N}.
\end{equation}
This leads to the following expression:
\begin{equation}
\widetilde P_{\cal N}
=\left(\ddq-({\cal N}-1)E(q)\right)
\left(\ddq-({\cal N}-2)E(q)\right)\cdots
\left(\ddq-E(q)\right)\ddq.
\end{equation}
Next we introduce a function $h(q)$ defined by the following;
\begin{equation}
h(q)=c_1\int^q_0 dq_1 e^{\int^{q_1}_0 E(q_2)dq_2} +c_2,
\label{eq:ehrel}
\end{equation}
where $c_{1,2}$ are constants.
The above is a general solution of the following differential equation:
\begin{equation}
h''(q)-E(q)h'(q)=0.
\label{eq:herel}
\end{equation}
We then find that $\widetilde P_{\cal N}$ may be written as follows:
\begin{equation}
\widetilde P_{\cal N}=
\left(h'\right)^{\cal N}\left(\frac{d}{dh}\right)^{\cal N}.
\end{equation}
We therefore arrive at the following simple expression:
\begin{equation}
P_{\cal N}=(-i)^{\cal N} U^{-1}
\left(h'\right)^{\cal N}\left(\frac{d}{dh}\right)^{\cal N}U.
\label{eq:pnsimple}
\end{equation}

The form (\ref{eq:pnsimple}) of $P_{\cal N}$ allows 
straightforward identification of the kernel ${\cal V}$:
The equations for its basis $\{\phi^-_1, \phi^-_2, \cdots, \phi^-_{\cal N}\}$;
\begin{equation}
P_{\cal N}\phi^-_n=0,
\end{equation}
can be simply solved as follows:
\begin{equation}
\phi^-_n=h^{n-1}U^{-1}  \quad (n=1,2,\cdots,{\cal N}).
\label{eq:phin1n}
\end{equation}

Next we need to find a Hamiltonian $H^-_{\cal N}$ that satisfy;
\begin{equation}
P_{\cal N}\hn\,\phi_n=0,
\label{eq:qscondition}
\end{equation}
for $n=1,2,\cdots,{\cal N}$.
By transforming the Hamiltonian $\hn$ by $U$ as 
\begin{equation}
\hn=U^{-1}\thn\, U,
\label{eq:hnthn}
\end{equation}
we find that the condition (\ref{eq:qscondition}) may be written as
follows:
\begin{equation}
\left(\frac{d}{dh}\right)^{\cal N} \thn \,h^{n-1}=0,
\label{eq:starcon}
\end{equation}
for $n=1,2,\cdots,{\cal N}$.
In the following, we will obtain $\thn$ that satisfy the above as a function of
$h$ and $d/dh$, noting that since $\thn$ contains second derivatives
with respect to $q$ it contains second derivatives with respect to $h$ as well.

Evidently, arbitrary constants are allowed in $\thn$ in Eq.(\ref{eq:starcon}).
The operators with first derivative with respect to $h$ allowed in $\thn$ are
given by the following:
\begin{equation}
\frac{d}{dh}, \quad
h\frac{d}{dh}, \quad
h^2\frac{d}{dh}-({\cal N}-1)h.
\label{eq:ddh1}
\end{equation}
It should be noted that in case of ${\cal N}=1$, the above is not the
complete list of such operators, since any operator of the form
$g(h)d/dh$ with an arbitrary function $g(h)$ is allowed.
Therefore, the following construction applies only for ${\cal N}\ge 2$.

All of the operators in the list (\ref{eq:ddh1}), 
when combined with constants, 
can be written in terms of \sltwo\  generators, 
whose representation on an ${\cal N}$-dimensional space spanned by
the basis $(1, h, h^2, \cdots, h^{{\cal N}-1})$ are the following:
\begin{eqnarray}
J^+ \equiv h^2\frac{d}{dh}-({\cal N}-1)h, \quad
J^0 \equiv h\frac{d}{dh}-\frac{{\cal N}-1}{2}, \quad
J^- \equiv \frac{d}{dh}.
\label{eq:first}
\end{eqnarray}
These generators satisfy the algebra,
\begin{equation}
[J^+, J^-]=-2J^0, \quad [J^\pm, J^0\,]=\mp J^\pm,
\end{equation}
and form the following Casimir operator:
\begin{equation}
\frac12\left(J^+J^-+J^-J^+\right)-(J^0)^2=-\frac14({\cal N}^2-1).
\label{eq:Casimir}
\end{equation}

Operators that contain second derivatives
with respect to $h$ that satisfy Eq.(\ref{eq:starcon}) are
only the following:
\begin{eqnarray}
\frac{d^2}{dh^2}&=&\left(J^-\right)^2, \\
h\frac{d^2}{dh^2}&=&J^0 J^- + \frac{{\cal N}-1}{2}J^-, \\
h^2\frac{d^2}{dh^2}&=&J^+ J^- + ({\cal N}-1)J^0+ \frac{({\cal N}-1)^2}{2},\\
h^3\frac{d^2}{dh^2}&-&({\cal N}-1)({\cal N}-2)h =
J^+J^0 -\frac{5-3{\cal N}}{2}J^+, \\
h^4\frac{d^2}{dh^2}&-&2({\cal N}-2)h^3\frac{d}{dh}
+({\cal N}-1)({\cal N}-2)h^2 = \left(J^+\right)^2.
\label{eq:second}
\end{eqnarray}
This time, the above list of the operators are complete only
for ${\cal N} \ge 3$, for which we restrict our construction.
From Eq.(\ref{eq:first})--(\ref{eq:second}), we find that 
the Hamiltonian $\thn$ that satisfy the \nfsusic\ condition (\ref{eq:starcon}) 
can be always written  in terms of the \sltwo\  
generators $J^{+,0,-}$ as follows:
\begin{equation}
\thn=-
\begin{array}[t]{c}
\displaystyle\sum_{i,j=+,0,-}\\[-3pt]
\scriptstyle j\ge i
\end{array} 
a_{ij}J^iJ^j + \sum_{i=+,0,-}b_iJ^i + C.
\label{eq:sl2desu}
\end{equation}
where $a_{ij}, b_i$ and $C$ are constants.

The Hamiltonian $\thn$ (\ref{eq:sl2desu}) 
is written in terms of $h$ as follows:
\begin{equation}
\thn = -P_4(h)\frac{d^2}{dh^2} + P_3(h)\frac{d}{dh}+P_2(h),
\label{eq:thnppp}
\end{equation}
where the coefficient functions $P_{4,3,2}(h)$ are;
\begin{eqnarray}
P_4(h)&=&a_{++}h^4 + a_{+0}h^3 +(a_{+-}+a_{00})h^2 + a_{0-}h + a_{--},
\label{eq:p4haaaa}\\[5pt]
P_3(h)&=&2({\cal N}-2)a_{++}h^3
+\left(\frac{3{\cal N}-5}{2}a_{+0}+b_+\right)h^2
\nonumber\\
&&\mbox{}+\left(({\cal N}-1)a_{+-}+({\cal N}-2)a_{00}+b_0\right) h
+\frac{{\cal N}-1}{2}a_{0-}+b_-,
\label{eq:p3aaaa}\\[5pt]
P_2(h)&=&-({\cal N}-1)({\cal N}-2)a_{++}h^2
-({\cal N}-1)\left(\frac{{\cal N}-1}{2}a_{+0}+b_+ \right)h\nonumber\\
&&\mbox{}-\frac{({\cal N}-1)^2}{4}a_{00}-\frac{{\cal N}-1}{2}b_0+C.
\label{eq:p2aaaa}
\end{eqnarray}
Transforming the Hamiltonian $\thn$ in Eq.(\ref{eq:thnppp}) back to 
the original Hamiltonian $\hn$ by the $U$-transformation (\ref{eq:hnthn})
and writing it in terms of the coordinate $q$, we find the following:
\begin{eqnarray}
\hn &=& 
-\frac{P_4(h)}{(h')^2}\frac{d^2}{dq^2}
+\left[ \frac{P_4(h)}{(h')^2}\left(-2W+\frac{h''}{h'}\right)
+\frac{P_3(h)}{h'}\right]\frac{d}{dq}\nonumber\\
&&\mbox{}-\frac{P_4(h)}{(h')^2}(W'+W^2)
+\frac{P_4(h)h''}{(h')^3}W +\frac{P_3(h)}{h'}W + P_2(h).
\end{eqnarray}
Comparing this with the regular form of the Hamiltonian (\ref{eq:Hfutu}), we
find the following requirement for $h(q)$ from the $d^2/dq^2$ term:
\begin{equation}
P_4(h)= \frac12(h')^2,\label{eq:for}
\label{eq:p4hprime}
\end{equation}
which is also a requirement for $E(q)$ through (\ref{eq:ehrel}).
Similarly, from the $d/dq$ term:
\begin{equation}
P_3(h)=h'\left(W-\frac{E}{2}\right).
\label{eq:lat}
\end{equation}
Finally, by the use of Eq.(\ref{eq:for})--(\ref{eq:lat}), the potential
$V^-_{\cal N}(q)$ is obtained as follows:
\begin{equation}
V^-_{\cal N}(q)=\frac12\left(W(q)^2 - W(q)'\right)+P_2(h(q)).
\label{eq:vminusp2}
\end{equation}

The other potential $V^+_{\cal N}(q)$ may be obtained from 
Eq.(\ref{eq:vpmdet}), or equivalently Eq.(\ref{eq:barVdef}):
Since Eq.(\ref{eq:pnsimple}) induces
\begin{equation}
P_{\cal N}=p^{\cal N} 
-i{\cal N}\ww p^{{\cal N}-1}+ [\,\mbox{terms with less powers of $p$}\,],
\end{equation}
we find that 
\begin{equation}
V^+_{\cal N}(q)=V^-_{\cal N}(q)+{\cal N}\ww'.
\label{eq:vplusres}
\end{equation}

We will now compare the \sltwo\ construction explained above
with the previous results on Type A \nfsusy\ \cite{AST,AST2}.
From Eq.(\ref{eq:p4haaaa})  we find the following identity:
\begin{equation}
\frac{d^5P_4}{dh^5}=0.
\end{equation}
On the other hand, using Eq.(\ref{eq:herel}) repeatedly on the expression
of $P_4(h)$ in Eq.(\ref{eq:p4hprime}), we find that 
\begin{equation}
0=\frac{d^5P_4}{dh^5}=
\frac{1}{h'^3}\left(\frac{d}{dq}-2E\right)
\left[(E'+E^2)''-E(E'+E^2)'\right],
\label{eq:genE}
\end{equation}
which is a generalization of one of the Type A conditions (\ref{eq:typeac}).
Eq.(\ref{eq:genE}) maybe rewritten as,
\begin{equation}
\frac{\left[(E'+E^2)''-E(E'+E^2)'\right]'}{(E'+E^2)''-E(E'+E^2)'}
=2E=2\frac{h''}{h'},
\end{equation}
which is integrated to yield that
\begin{equation}
(E'+E^2)''-E(E'+E^2)'=\beta_1 \left(h'\right)^2,
\label{eq:beta1E}
\end{equation}
where $\beta_1$ is an integration constant.
The above is further integrated to yield the following:
\begin{equation}
E'+E^2=\frac12\beta_1 h^2 + \beta_2 h +\beta_3,
\end{equation}
where $\beta_{2,3}$ are integration constants.
Since we have 
\begin{equation}
\frac{d^2P_4}{dh^2}=E'+E^2,
\end{equation}
from Eq.(\ref{eq:p4hprime}), we find the following identification of
the constants:
\begin{equation}
a_{++}=\frac{\beta_1}{4!}, \quad
a_{+0}=\frac{\beta_2}{3!}.
\label{eq:i1}
\end{equation}

Similarly to Eq.(\ref{eq:genE}), we can obtain the following:
\begin{equation}
0=-\frac{{\cal N}-2}{2}\frac{d^4P_4}{dh^4}+\frac{d^3P_3}{dh^3}
=\frac{1}{h'^2}\left[(\wwq'+E\wwq)''-E(\wwq'+E\wwq)'\right],
\end{equation}
which is exactly one of the Type A conditions (\ref{eq:typeab}).
This equation is integrated to give the following:
\begin{equation}
\wwq'+E\wwq=\beta_4 h + \beta_5,
\end{equation}
where $\beta_{4,5}$ are integration constants.
By comparing the above and 
\begin{equation}
\frac{dP_3}{dh}=(\wwq'+E\wwq)+\frac{{\cal N}-2}{2}(E'+E^2),
\end{equation}
we find the following identification:
\begin{equation}
b_+=\frac12\beta_4 - \frac1{12}\beta_2.
\label{eq:i2}
\end{equation}

Using Eqs.(\ref{eq:i1}) and (\ref{eq:i2}), we can obtain the following
expression of $P_2(h)$ in terms of $E(q)$ and $W(q)$:
\begin{equation}
P_2(h)=-\frac{{\cal N}-1}{2}
\left[\frac{{\cal N}-2}{6}\left(E'+E^2\right) + \left(\wwq'+E\wwq\right)\right]
+ \mbox{constants},
\label{eq:P2final}
\end{equation}
which, together with  Eqs.({\ref{eq:vminusp2}) and (\ref{eq:vplusres}),
reproduces the potentials $V^\pm_{\cal N}(q)$ in Eq.(\ref{eq:typeaa}).

In summary, we have shown that \sltwo\ construction yields
Type A \nfsusy,
which is defined by the form of the supercharge (\ref{eq:typeapn}).
We find that following set of conditions is
sufficient for ${\cal N}=1,2$
and is necessary and sufficient for ${\cal N}\ge 3$:
\begin{eqnarray}
&&V^{\pm}_{\cal N}=\frac{1}{2}\left(W^2 \pm W'\right)\nonumber\\
&&\hspace{30pt} \mbox{}+\frac{{\cal N}-1}{2}
\left[-EW+\frac{2{\cal N}-1}{6}E^2-\frac{{\cal N}+1}{6}E'
\pm\left(W'-\frac{\cal N}{2}E'\right)\right],
\label{eq:typeaag}\\
&&(\wwq'+E\wwq)''-E(\wwq'+E\wwq)'=0,
\label{eq:typeabg}\\
&&\left(\frac{d}{dq}-2E\right)
\left[(E'+E^2)''-E(E'+E^2)'\right]=0,
\label{eq:typeacg}
\end{eqnarray}
in place of Eqs.(\ref{eq:typeaa})--(\ref{eq:typeac}).
We note that one may conversely derive the \sltwo\ form from the above,
by solving these equations and defining
$P_{4,3,2}(q)$ as in Eqs.(\ref{eq:p4hprime}), (\ref{eq:lat}) and
(\ref{eq:P2final}), respectively.

We have found above that \sltwo\ construction contains the original conditions
Eqs.(\ref{eq:typeaa}) --(\ref{eq:typeac}), but gives a less-restrictive
condition (\ref{eq:typeacg}).
This is explained by the fact that original conditions were obtained 
by induction in ${\cal N}$.
In such an induction, 
we implicitly
assumed that $E(q)$ and $W(q)$ are independent from ${\cal N}$.
In our \sltwo\ construction, ${\cal N}$-independence of $E(q)$ 
implies ${\cal N}$-independence of $h(q)$ (assuming
that the coefficients $c_1$ and $c_2$ in Eq.(\ref{eq:ehrel})
are ${\cal N}$-independent as well)
and thus of $P_4(q)$ through (\ref{eq:for}),
which in turn means ${\cal N}$-independence of all the $a$-coefficients
in Eq.(\ref{eq:p4haaaa}).
Further, ${\cal N}$-independence of $W(q)$ 
implies ${\cal N}$-independence of all four coefficients of
$h^{3,\cdots,0}$ terms in $P_3(q)$ in Eq.(\ref{eq:p3aaaa}).
The ${\cal N}$-independence of coefficient of $h^3$ term
implies that
\begin{equation}
a_{++}=0,
\label{eq:appzero}
\end{equation}
while the latter three coefficients may be made ${\cal N}$-independent
by appropriate choice of the $b$-coefficients.
Using our identification (\ref{eq:i1}), we find that
the above corresponds to $\beta_1=0$, which, as seen in Eq.(\ref{eq:beta1E}),
reproduces the the Type A condition (\ref{eq:typeac}).
\footnote{Recently we have proven the conditions 
(\ref{eq:typeaag})--(\ref{eq:typeacg}) by a direct calculation.
This proof will be published in near future.}

\section{Specific Examples} 
In this section, we illustrate the correspondence between some specific
examples of the type A potentials, some of 
which appear in Ref.\cite{AST}, and some of the \sltwo\
models mentioned in Ref.\cite{Tur}.

In Ref.\cite{Tur}, for a quasi-solvable potential $V(q)$  and 
the Hamiltonian,
\begin{equation}
H = -\frac12\frac{d^2}{dx^2} + \frac12V(q),
\end{equation}
the wave function can be written in the following form:
\begin{equation}
\psi(q) = \varphi(z(q))\  e^{-g(q)}.
\label{eqn:turwf}
\end{equation}
Therefore, by comparing (\ref{eq:phin1n}})
with (\ref{eqn:turwf}), we suppose the
general relations among the two formalisms are 
\begin{equation}
h(q) = z(q),
\label{eqn:taioua}
\end{equation}
\begin{equation}
W(q) = \frac{d}{dq}g(q).
\label{eqn:taioub}
\end{equation}
We note that in the following we will write down only
the potential $V_{\cal N}^{-}(q)$.
The other potential $V_{\cal N}^{+}(q)$
can be constructed easily using the relation (\ref{eq:vplusres}).
Throughout this analysis, we neglect constant terms in the potentials.

\subsection{Quadratic Type}
First, we consider the case of $E=0$, which is a trivial
solution of Eq.(\ref{eq:typeacg}). In this case, the
following quadratic $W(q)$ and quartic potential $V_{\cal N}^{-}(q)$ are
obtained from Eqs.(\ref{eq:typeaag}) and (\ref{eq:typeabg}):
\begin{equation}
W(q)=C_1 q^2 + C_2 q + C_3,
\end{equation}
\begin{equation}
V_{\cal N}^{-}(q)= 
\frac{C_1^2}{2} q^4+ C_1 C_2 q^3+\left(\frac{C_2^2}{2}+  C_1 C_3 \right)q^2
        + (  C_2 C_3 -  {\cal N}C_1 )q.  
\end{equation}
This type of quartic potential is not
mentioned in Ref.\cite{Tur}.\footnote{This may be because
this potential makes sense only perturbatively:
The wavefunction is normalizable at any finite order of
the perturbation theory, but not at full order.
Therefore, the obtained energy eigenvalues represent only
the perturbative part \cite{AKOSW,AST2}.}
However, we see easily
that this example is constructed from \sltwo\ generators.  
By transforming $H_{\cal N}^-$ by $U$ as (\ref{eq:hnthn}) and setting
$h(q)=q$, which is consistent with $E=0$ and Eq.(\ref{eq:herel}), we get
\begin{equation}
\thn = -\frac12\frac{d^2}{dh^2} +(C_1 h^2+C_2 h +C_3) \frac{d}{dh}
-  C_1 ({\cal N}-1)h.
\end{equation}
This can be put into the bilinear and linear form of the \sltwo\ generators;
\begin{equation}
\tilde{H}_{\cal N}^- = -\frac12(J^-)^2 + C_1 J^+ + C_2 J^0 + C_3 J^-.
\end{equation}
This reproduces the form noted in Footnote 12 in Ref.\cite{AKOSW2}
with suitable choice of $C_i$.

\subsection{Exponential and Periodic Type}
Next, we take $E=E_0$\ (a non-zero constant), which also trivially
satisfies Eq.(\ref{eq:typeacg}).
In this case, we find the following 
from Eqs.(\ref{eq:typeaag}) and (\ref{eq:typeabg}):
\begin{equation}
W(q) = C_1 e^{E_0 q}+ C_2 e^{-E_0 q}+C_3,
\label{eqn:expw}
\end{equation}
\begin{equation}
V_{\cal N}^-(q) = \frac{C_1^2}{2} e^{2E_0q}+ 
\frac12C_1 \{2C_3 -E_0(2{\cal N}-1)\}e^{E_0 q}
+ \frac12C_2\left(2 C_3 + E_0 \right)e^{-E_0 q}
+\frac{C_2^2}{2} e^{-2E_0q}.
\label{eqn:expva}
\end{equation}
When $E_0$ is chosen to be real, 
this exponential potential corresponds to the potential (I) in Ref.\cite{Tur}.
Furthermore, 
when $E_0$ is chosen to be complex and coefficients are appropriately
chosen so that potential is real, the above potential reproduces
the potential (X) in Ref.\cite{Tur}.
The potential and the wave function there are as follows:
\begin{equation}
V(q) = a^2 e^{-2 \alpha q} -a\{2b+ \alpha (2{\cal N}-1) \}e^{- \alpha q}+c(2b - \alpha)e^{\alpha q}+c^2 e^{2 \alpha q},
\label{eqn:expvt}
\end{equation}
\begin{equation}
\varphi(q) = A_0 e^{- \alpha ({\cal N}-1) q} + A_1 e^{- \alpha ({\cal N}-2) q}+ \ldots +A_{{\cal N}-1},
\end{equation}
\begin{equation}
g(q) = \frac{a}{\alpha} e^{- \alpha q }
+b q +  \frac{c}{\alpha} e^{\alpha q} ,
\label{eqn:expg}
\end{equation}
\begin{equation}
z(q) = e^{- \alpha q}.
\label{eqn:expz}
\end{equation}
Therefore, from Eq.(\ref{eq:herel}), (\ref{eqn:taioua}) and (\ref{eqn:expz}), 
we obtain the relation;
\begin{equation} 
E_0 = - \alpha.
\end{equation}
In addition, 
the relations of the other parameters are determined 
from Eqs.(\ref{eqn:taioub}), (\ref{eqn:expw}) and (\ref{eqn:expg});
\begin{equation} 
C_1 = -a,\
C_2 = c,\
C_3 = b,
\end{equation}
and then the potential (\ref{eqn:expva}) equals to (\ref{eqn:expvt}).

\subsection{Cubic Type}
Furthermore, we consider a solution of (\ref{eq:typeacg}),
$E=1/q$.
In this case,
we get the following $W(q)$ and the potential $V_{\cal N}^-(q)$
in the same way as above:
\begin{equation}
W(q) = C_1 q^3 + C_2 q + C_3 \frac{1}{q},
\label{eqn:cubw}
\end{equation}
\begin{equation} 
V_{\cal N}^-(q) = \frac{C_1^2}{2} q^6 + C_1 C_2 q^4 
      + \frac12\{C_2^2-(4 {\cal N}-1-2 C_3) C_1\} q^2
+\frac{C_3(C_3 + 1)}{2}\frac{1}{q^2}. 
\label{eqn:cubva}
\end{equation}
This sextic potential is equivalent to the potentials 
(VI) and (VII) in Ref.\cite{Tur}.
The potentials and the wave functions\footnote{The potential and the 
wave function
are originally given as an example of the spherically symmetric
quasi-solvable models. As we are discussing one-dimensional potentials 
here, we set $d=1$ and $l=0$.} 
are given by the following:
\begin{equation}
V(q) = a^2 q^6 + 2 a b q^4 +\{ b^2 - (4 {\cal N}-1-2 c)a \}q^2 
+ c(c+1)\frac{1}{q^2},
\label{eqn:cubvt}
\end{equation}
\begin{equation}
\varphi (q) = A_0 (q^2)^{{\cal N}-1} + A_1 (q^2)^{{\cal N}-2}+ \ldots
+A_{{\cal N}-1}, 
\end{equation}
\begin{equation}
g(q) =  \frac{a}{4}q^4 + \frac{b}{2}q^2 + c \ln{q}, 
\label{eqn:cubg}
\end{equation}
\begin{equation}
z(q) = q^2.
\label{eqn:cubz}
\end{equation}
Therefore, from (\ref{eqn:taioub}), (\ref{eqn:cubw}) and (\ref{eqn:cubg}),
the relations of the parameters are determined as follows:
\begin{equation} 
C_1 = a,\
C_2 = b,\
C_3 = c.
\end{equation}
Under these relations, the potential (\ref{eqn:cubva}) equals to the potential
(\ref{eqn:cubvt}). Note that $E = 1/q$ is consistent with
Eqs.(\ref{eq:herel}), (\ref{eqn:taioua}) and (\ref{eqn:cubz}).

\subsection{Hyperbolic Type}
At the end of this section, we consider another solution 
of Eq.(\ref{eq:typeacg}),
\begin{equation}
E= \alpha \left( \frac{1}{\cosh{\alpha q}\sinh{\alpha q}} - 2
\tanh{\alpha q} \right).
\label{eqn:hype}
\end{equation}
In this case, the following $W(q)$ and the potential $V_{\cal N}^-(q)$
are obtained:
\begin{equation}
W(q) =  \frac{C_1}{\cosh{\alpha q} \sinh{\alpha q}} + C_2 \cosh{\alpha
q} \sinh{\alpha q}+ C_3 \tanh{\alpha q}, 
\label{eqn:hypw}
\end{equation}
\begin{eqnarray}
V_{\cal N}^- (q) &=& \frac{C_2^2}{2}\cosh^4{\alpha q}
- C_2 \left( \frac{C_2}{2} - C_3 + \alpha \right)
 \cosh^2{\alpha q} \nonumber \\
&& \mbox{}-\left[\frac{C_3^2}{2} - C_1 C_3 - \alpha (2{\cal N}-1) 
\left\{ C_1 - ({\cal N}-1) \alpha \right\} 
+\frac{C_3}{2}\alpha (4{\cal N}-3) \right] \frac{1}{\cosh^2{\alpha q}}  
\nonumber \\        
&& \mbox{}+\frac{C_1}{2}(C_1+\alpha) 
\frac{1}{\cosh^2{\alpha q}\sinh^2{\alpha q}}.    
\label{eqn:hypva}
\end{eqnarray}
This type of hyperbolic potential corresponds to the potential (IV) 
in Ref.\cite{Tur}.
The potential and wave function are 
\begin{equation}
V(q) = c^2 \cosh^4{\alpha q} -c(c+2 \alpha -2 a)\cosh^2{\alpha q}
-\{a(a + \alpha)+ \alpha k (\alpha k + \alpha +2 a) \}\cosh^{-2}{\alpha q },
\label{eqn:hypvt}
\end{equation}
\begin{equation}
\varphi(q) = A_k \tanh^k{\alpha q} + A_{k-1} \tanh^{k-1}{\alpha q} +
\ldots + A_0,
\end{equation}
\begin{equation}
g(q) = \displaystyle \frac{c}{4 \alpha}  \cosh{2 \alpha q} + \frac{a}{\alpha} \ln{\cosh{\alpha q}},
\label{eqn:hypg}
\end{equation}
\begin{equation}
z(q) = \cosh^{-2}{\alpha q},
\label{eqn:hypz}
\end{equation}
where ${\cal N}= [\,k/2\,] + 1$.
Also, $A_{i={\rm odd}}=0$ for even $k$ 
and $A_{i={\rm even}}=0$ for odd $k$.
The solution (\ref{eqn:hype}) is consistent with Eqs.(\ref{eq:herel}),
(\ref{eqn:taioua}) and (\ref{eqn:hypz}), 
and we get the following relations 
from Eqs.(\ref{eqn:taioub}), (\ref{eqn:hypw}) and (\ref{eqn:hypg}) for 
even $k$;
\begin{equation}
C_1 = 0, \quad C_2 = c,\quad C_3 = a.
\end{equation}
When these relations hold, the potential (\ref{eqn:hypva}) 
equals to the potential (\ref{eqn:hypvt}). 
For odd $k$, we modify Eq.(\ref{eqn:taioub}) as,
\begin{equation}
W(q)=\frac{d}{dq}g(q)-\frac{\alpha}{\cosh\alpha q \sinh\alpha q},
\end{equation}
to find that the relations,
\begin{equation}
C_1 = -\alpha, \quad C_2 = c,\quad C_3 = a,
\end{equation}
reproduce the potential and the wave function.

We note that some other models are also known, notably ones with
$E=-3/q$ and $E=\alpha/\tan(\alpha q)$.
All of those models and some other new models we have found recently
will be discussed in a separate literature.

\section{Summary}
In this paper we have shown that by quasi-solvability considerations
\sltwo\ emerges naturally and \nfsusy\ can be constructed by the use
of \sltwo\ generators.
This is done by first constructing the 
kernel space of the supercharge $P_{\cal N}$ of the Type A form
(\ref{eq:typeapn}) and then showing that 
the Hamiltonian that leave the kernel invariant
is always written in terms of \sltwo\ generators as Eq.(\ref{eq:sl2desu}).
By requiring that this Hamiltonian induces the canonical form
(\ref{eq:Hfutu}), we have obtained a condition (\ref{eq:typeacg}) on $E(q)$,
a relation (\ref{eq:typeabg}) between $E(q)$ and $W(q)$,
and have found a expression (\ref{eq:typeaag})
of the potentials $V_{\cal N}^\pm(q)$ in terms of $E(q)$ and $W(q)$.
These equations are more general than the previously obtained
equations (\ref{eq:typeaa})--(\ref{eq:typeac}) and are necessary and 
sufficient for Type A \nfsusy.  
We have also examined explicitly known models in the
light of the \sltwo\ construction.

\section*{Acknowledgments}
H. Aoyama's work was supported in part by the Grant-in-Aid for
Scientific Research No.10640259.
T. Tanaka's work was supported in part by a JSPS research fellowship.

\end{document}